\documentclass[preprint, review, 12pt]{elsarticle}

\usepackage{amssymb}
\usepackage{amsmath}
\usepackage{lineno}
\usepackage{mhchem}
\usepackage{booktabs}
\usepackage{multirow}
\usepackage{graphicx}
\usepackage{booktabs}
\usepackage{array}
\usepackage{tabularx}
\usepackage{chemformula}   
\usepackage{array} 
\usepackage{float}

\newcolumntype{P}[1]{>{\centering\arraybackslash}p{#1}}

\journal{Optics and Laser in Engineering}

\begin{document}

\begin{frontmatter}



\title{Multipolar Origin and Active Control of High-Q Quasi-BIC Fano Resonances in Dielectric Metasurfaces for Sensing Applications}


\author[lev1]{Soikot Sarkar} 
\author[lev1]{Ahmed Zubair\corref{cor1}}

\ead{ahmedzubair@eee.buet.ac.bd}
\cortext[cor1]{Corresponding Author.}
\affiliation[lev1]{organization={Department of Electrical and Electronic Engineering, Bangladesh University of Engineering and Technology},
            city={Dhaka},
            postcode={1205},
            country={Bangladesh}}

\begin{abstract}
We designed an ingenious all-dielectric metasurface, employing cuboid structures patterned with bow-tie-shaped nanoholes, exhibiting multiple Fano resonances induced by quasi-bound states in the continuum (quasi-BICs) through structural asymmetry. Among them, several resonant modes demonstrated high quality factors in the range of $10^3$–$10^4$, along with near-unity modulation depth and strong spectral contrast. The optical responses were analyzed utilizing the finite-difference time-domain (FDTD) method, with Fano profiles fitted to theoretical models and the BIC-governed modes validated via the squared inverse ratio law. Furthermore, multipolar decomposition and electromagnetic spatial field profile revealed the origins of the resonance, while LC circuit modeling provided additional physical insight into the Fano profiles. The proposed metasurface also exhibited strong polarization dependence, indicating its potential for active optical switching. Finally, refractive index sensing performance, including the detection of \textit{Vibrio cholerae}, reached a sensitivity of 342 nm/RIU and a figure of merit of 217.14 RIU\textsuperscript{-1}. Advancing the control of high-Q quasi-BIC Fano resonances, this study highlights Fano resonators' potential for refractive index sensing and active switching.
\end{abstract}





\end{frontmatter}



\section{Introduction}
Metasurfaces with engineered strong light–matter interactions enable advanced and multifunctional control over optical wavefronts, including amplitude, phase, polarization, and dispersion~\cite{Shaltout2019, Pertsch2023, Dip2025JMCC, Sarkar2024, Nakti2023}. Their compactness, high design flexibility, and ability to integrate with existing platforms make them ideal candidates for next-generation technologies in flat optics, compact imaging, on-chip photonic circuits, and broadband optical communication~\cite{Kuznetsov2024, Koshelev2019, Islam2025}. By harnessing sharp spectral features arising from resonant effects, such as Fano resonances and bound states in the continuum (BIC), these planar photonic structures provide exceptional capability for light manipulation at subwavelength scales~\cite{Chen2024, Kang2023, Yi2025, Jafari2024}. This ability depends critically on the material’s inherent optical characteristics and the precise engineering of resonant nanostructures. Metasurfaces are broadly categorized into plasmonic and all-dielectric types, each distinguished by their resonance mechanisms and electromagnetic response~\cite{Lv2024}. Plasmonic metasurfaces, typically composed of noble metals or graphene, rely on localized surface plasmon resonances to achieve strong near-field confinement in the visible to near-infrared range~\cite{Yang2024, Sarker2021}. However, their practical utility is constrained by inherent ohmic losses, which limit Q factors and degrade performance in applications requiring high spectral selectivity~\cite{Liang2024}. Notably, all-dielectric metasurfaces constructed from high refractive index materials, such as silicon, germanium, gallium phosphide, support low-loss Mie resonances and enable the excitation of high-Q modes, including BICs, through careful control of geometric and symmetry parameters. These structures exhibit minimal absorption losses, enhanced field confinement, and are compatible with CMOS fabrication processes, making them particularly attractive for scalable and low-loss photonic integration~\cite{Wang2023, Wang2021, Song2023}.\\

Among the various multipolar excitations in nanophotonics, the toroidal dipole (TD), first introduced by Zeldovich in 1957, has emerged as a distinctive mode arising from closed-loop displacement currents, representing a unique excitation in all-dielectric metasurfaces with weak radiation and behavior markedly different from conventional electric (ED) and magnetic dipole (MD) responses~\cite{Xu2021}. Although TD modes were previously overlooked due to masking by dominant multipoles, recent studies have demonstrated that their excitation can be selectively enhanced through structural asymmetry, near-field coupling, and engineered high-index dielectric arrays~\cite{Li2022, Li2021}. These configurations support a high Q factor with suppressed radiative loss, making them promising for applications in sensing and nonlinear optics. However, the relatively weak near-field confinement of TD modes highlights further optimization in geometry and material design to maximize their photonic functionality.\\

Ultra-narrow optical resonances with exceptionally high Q factors and minimal losses play a vital role in modern nanophotonics, enabling enhanced light–matter interactions essential for applications such as high-sensitivity sensors, nonlinear optics, slow-light devices, narrowband absorbers, and spectral filters~\cite{Sarkar2024, Sarker2024_PCCP, Xie2023, Chen2025, Feng2023, Ye2022}. A common route to achieving such resonances is through Fano resonance, resulting from interference between radiative and non-radiative modes, producing sharp asymmetric line shapes~\cite{Wang2024, Wang2024_AO}. This effect becomes particularly pronounced when combined with BICs, arising from non-radiating localized states embedded within the continuous radiation spectrum~\cite{Yang2021}. Although ideal BICs possess infinite Q factors and zero linewidth, practical implementation is achieved through quasi-BICs, where slight structural asymmetry induces weak coupling to free-space radiation, resulting in finite but extraordinarily high Q factors~\cite{Bhowmik2024, Liu2024_RSC, Dong2025, Zhao2024}. Such mechanisms have been widely utilized to excite high-Q TD Fano resonances in all-dielectric metasurfaces~\cite{Sun2024}. Pang \textit{et al.} introduced square defects into \ce{Si} tetramer arrays to realize polarization-independent quasi-BICs~\cite{Pang2024}. Sun \textit{et al.} employed a nanorod–ring unit with broken symmetry, enabling tunable TD resonances by varying the nanorod length~\cite{Sun2024}. Li \textit{et al.} achieved BIC-governed high-Q resonances in a \ce{Si}-based hollow metasurface by incorporating asymmetric air holes~\cite{Li2021}. Additionally, Lv \textit{et al.} investigated quadrupolar quasi-BICs in GaP nanodisk arrays activated via lateral displacement~\cite{Lv2024}. However, beyond geometric asymmetry, this effect can also be realized through material asymmetry. Yu \textit{et al.} demonstrated dual-band, polarization-insensitive TD modes using permittivity-asymmetric \ce{Si}–\ce{InAs} cuboid tetramers~\cite{Yu2022}. Although \ce{GaP} exhibits a high refractive index comparable to \ce{Si}, comprehensive investigations of \ce{GaP}-based metasurfaces, especially regarding BIC-governed Fano resonances, remain insufficient. Moreover, hollow-structured \ce{GaP} metasurfaces tailored for such resonance phenomena have yet to be thoroughly explored.\\\\

In this paper, we present an all-dielectric metasurface exhibiting multiple Fano resonances in the near-infrared (NIR) regime, predominantly governed by MD and TD excitations. The incorporation of structural asymmetry—achieved by modifying both the dimension and orientation of a single triangular nanohole of bow-tie-shaped nanoholes—introduced quasi-BIC-supported Fano resonances, resulting in additional pronounced Fano resonances. These resonances exhibited both exceptionally high quality factors and modulation depths. The Fano profiles for both symmetry and asymmetry configurations were fitted using the theoretical model, while the BIC mode was validated by calculating the squared inverse ratio law. Additionally, the Fano resonances were interpreted using an equivalent LC circuit model to elucidate the underlying resonance mechanism. To further analyze the origin of resonance, the contributions of various multipolar moments were investigated through spatial distributions of the electric and magnetic fields. The impact of the polarization angle on optical performance was investigated, demonstrating potential for being utilized as an optical switch. Moreover, the impact of the structural parameters was also investigated. Finally, the sensing capability of the structure was assessed by changing the ambient refractive index to simulate the presence of Vibrio cholerae. This all-dielectric metasurface, with lower complexity in design, offers promising prospects for advanced applications in RI sensing and active photonic switching.

\section{Structure Design and Methodology}
\begin{figure*}[htbp]
    \centering
    \includegraphics[width=1\linewidth]{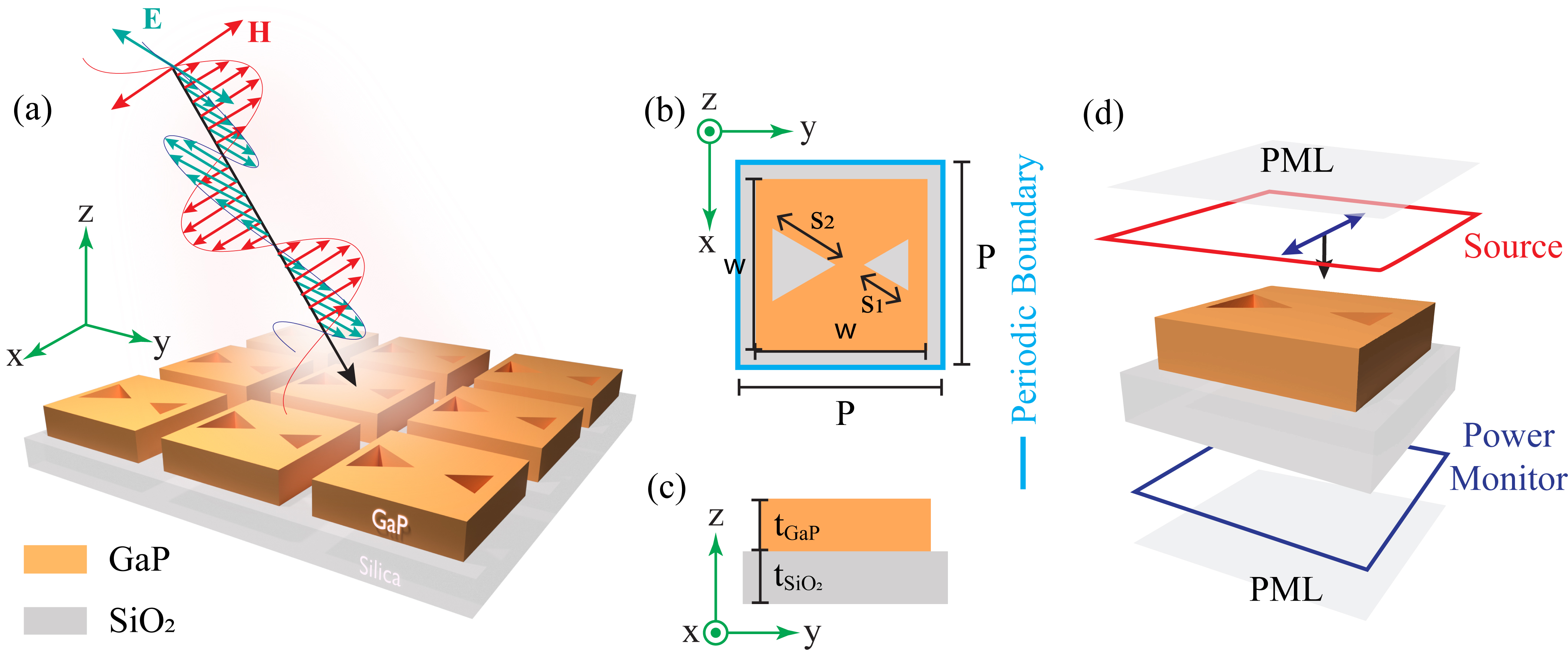}
    \caption{(a) 3D schematic illustration of the proposed all-dielectric bow-tie etched metasurface, (b) x-y, and (c) x-z plane view of the structure. The optimized structural parameters: P = $650~\mathrm{nm}$, w = $550~\mathrm{nm}$, t\textsubscript{SiO\textsubscript{2}} = $200~\mathrm{nm}$, t\textsubscript{GaP} = $150~\mathrm{nm}$, and $\delta$ = s\textsubscript{2} - s\textsubscript{1}. (d) Optical simulation configuration of the proposed structure.}
    \label{fig: Structure}
\end{figure*}

Figure~\ref{fig: Structure}(a) depicts our proposed all-dielectric metasurface, comprising a two-dimensional periodic array of \ce{GaP} nanocuboids with square cross-section and bow-tie-shaped nanoholes, deposited on a \ce{SiO2} substrate along the x and y axes with a periodicity, P $=650~\mathrm{nm}$.  
The thickness of the substrate was considered to be t\textsubscript{\ce{SiO2}} $=200~\mathrm{nm}$. Moreover, Figs.~\ref{fig: Structure}(b) and (c) present the cross-sectional views along the xy- and yz-planes, respectively. The side of the square-shaped cross-section of \ce{GaP} nanocuboids was w $=550~\mathrm{nm}$, while the thickness was t\textsubscript{\ce{GaP}} $=150~\mathrm{nm}$. The bow-tie-shaped nanoholes were comprised of two equilateral triangles with side lengths s\textsubscript{1}, and s\textsubscript{2}, separated from center to center with a distance of $275~\mathrm{nm}$. To facilitate the BIC, an asymmetry parameter $\delta$ was introduced, defined as $\delta =$ s\textsubscript{2} - s\textsubscript{1}. The complex refractive index of \ce{GaP} and \ce{SiO2} were adopted from Bond \textit{et al.}~\cite{Bond1965} and Palik \textit{et al.}~\cite{Palik1998}, respectively [see \textcolor{blue}{Supplementary Material} for details]. \\
We employed the finite-difference time-domain (FDTD) approach in Ansys Lumerical to quantitatively investigate the optical properties of our proposed structure. The simulation configuration is illustrated in Fig.~\ref{fig: Structure}(d). Since our proposed metasurface exhibited periodicity along the x and y axes, we employed periodic boundary conditions to reduce computational space and simulation time while applying 12-layer perfectly matched layer (PML) boundary conditions along both directions of the z-axis to eliminate parasitic reflections from the structure. An additional mesh with dimension $4~\mathrm{nm} \times 4~\mathrm{nm} \times 4~\mathrm{nm}$ was incorporated to simulate precisely. The simulation was conducted by considering the structure immersed in a liquid medium, such as water, with a refractive index of $1.33$ at the temperature of $300$ K. A CW-normalized plane wave source was incident from the top of the structure with a wavelength ranging from $900~\mathrm{nm}$ to $1600~\mathrm{nm}$. We employed a frequency-domain field and power monitor beneath the structure to quantify transmittance $T(\lambda)$, which is the ratio of transmitted power, $P_T(\lambda)$, to the incident power, $P_I(\lambda)$, expressed as follows~\cite{Sarkar2024}, 
\begin{equation}
    T(\lambda)=\frac{P_T(\lambda)}{P_I(\lambda)}.
\end{equation}

\section{Results and Discussion}
\subsection{Dimensional Asymmetry}
\begin{figure*}[htbp]
    \centering
    \includegraphics[width=1\linewidth]{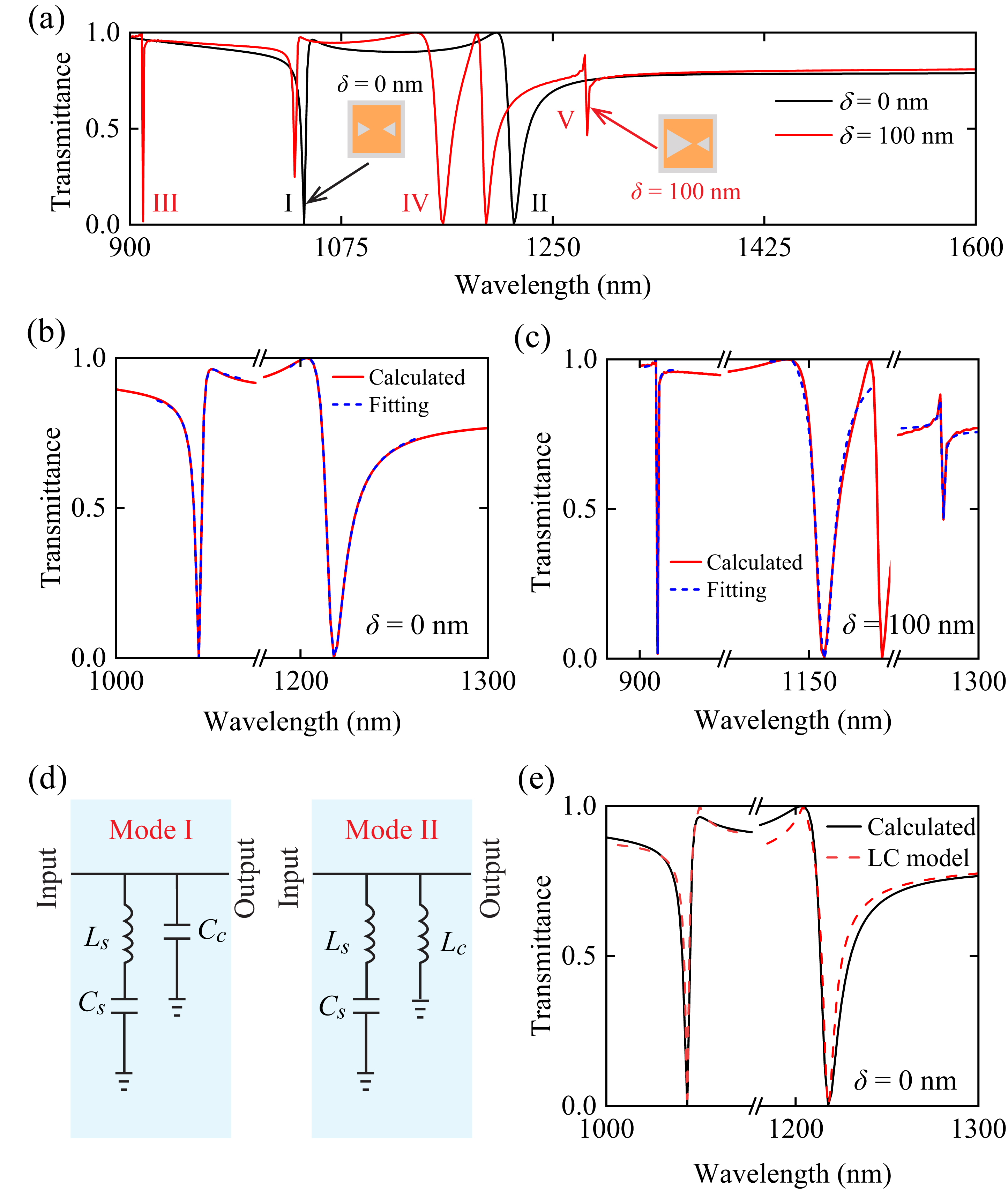}
    \caption{(a) Transmittance spectra of our proposed structure for $\delta = 0~\mathrm{nm}$, and $100~\mathrm{nm}$. Plot of FDTD calculated and fitted graph for (b) $\delta = 0~\mathrm{nm}$ near $\lambda = 1044~\mathrm{nm}$, and $1217~\mathrm{nm}$ (c) $\delta = 100~\mathrm{nm}$ near $\lambda = 911~\mathrm{nm}$, $1159~\mathrm{nm}$, and $1278~\mathrm{nm}$ under TM-polarized incident light. (d) Schematics of the LC circuit models for Mode I and II ($\delta = 0~\mathrm{nm}$). (e) FDTD calculated and fitted curve with LC circuit for Mode I and II for $\delta = 0~\mathrm{nm}$.}
    \label{fig: diff 0 100 & fit}
\end{figure*}

The asymmetry in dimensions of the bow-tie-shaped nanoholes allowed our proposed structure to exhibit resonances governed by BIC. Figure~\ref{fig: diff 0 100 & fit}(a) illustrates the transmittance spectra of the structure for $\delta = 0~\mathrm{nm}$ (s\textsubscript{1} = s\textsubscript{2} = $130~\mathrm{nm}$), and $\delta = 100~\mathrm{nm}$ (s\textsubscript{1} = $130~\mathrm{nm}$, s\textsubscript{2} = $230~\mathrm{nm}$). 
When $\delta = 0~\mathrm{nm}$, the structure maintained a distinct in-plane symmetry, resulting in two sharp Fano profiles at $\lambda = 1044~\mathrm{nm}$ (Mode I), and $1217~\mathrm{nm}$ (Mode II). The structure also supported symmetry-protected BICs with theoretically infinite Q factors but remained undetectable in the transmittance spectrum. However, the in-plane symmetry was disrupted when we considered $\delta = 100~\mathrm{nm}$,  allowing the non-radiative bound states to couple with the free space continuum and access the radiation channel~\cite{Li2021}. Consequently, the radiation channel transformed into a resonant mode,  allowing energy leakage and outward radiation, which led to the transformation of symmetry-protected BICs into quasi-BICs~\cite{Lv2024}. Therefore, we observed three additional Fano profile resonance at $\lambda = 911~\mathrm{nm}$ (Mode III), $1159~\mathrm{nm}$ (Mode IV), and $1278~\mathrm{nm}$ (Mode V) governed by quasi-BICs. Meanwhile, the Fano profile (Mode I and II) of this structure 
experienced a blue shift at $\lambda = 1036~\mathrm{nm}$, and $1195~\mathrm{nm}$, respectively.\\

The performance of the metasurface in optical applications can be evaluated based on the modulation depth, spectral contrast ratio, and Q-factor of the Fano resonance. The modulation depth is defined by~\cite{Yu2022},
\begin{equation}
    Modulation~depth = \frac{T_{peak}-T_{dip}}{T_{peak}}\times100\%
\end{equation}
where $T_{peak}$, and $T_{dip}$ are the maximum and minimum values of the transmittance spectrum, respectively. Moreover, we calculated the spectral contrast ratio by utilizing these two parameters,
\begin{equation}
    Spectral~contrast~ratio= \frac{T_{peak}-T_{dip}}{T_{peak}+T_{dip}}\times100\%.
\end{equation}
To observe the transmittance spectra qualitatively, the Fano model can be expressed as,
\begin{equation}
\label{fano model}
    T_{Fano} = \left| a_1 + i a_2 + \frac{b}{\omega - \omega_{s_\circ} + i \gamma} \right|^2,
\end{equation}
where $a_1$, $a_2$, and $b$ are the real number constants, $\omega_{s_\circ}$, and $\omega$ are the resonance and angular frequency, and $\gamma$ is the overall damping loss. The radiative Q-factor, $Q_{rad}$, can be defined as $Q_{rad} = \omega_{s_\circ/2}\gamma$. We employed Eq.~\ref{fano model} to fit the calculated modes I and II for $\delta = 0~\mathrm{nm}$ and III to V for $\delta = 100~\mathrm{nm}$. Figures~\ref{fig: diff 0 100 & fit} (b) and (c) present the calculated and fitted Fano profiles (Mode I and II) for $\delta = 0~\mathrm{nm}$ and (Mode III to V) for $100~\mathrm{nm}$, demonstrating that the fitted curve closely aligns with the calculated data. Table~\ref{tab: Quality factor} summarizes the extracted $Q_{rad}$ values from these fittings, modulation depth, and spectral contrast ratio for modes I, II ($\delta=0~\mathrm{nm}$) and I-V ($\delta = 100~\mathrm{nm})$. \\
\begin{table*}[]
\centering
\caption{Extracted values of quality factor, modulation depth, and spectral contrast ratio for modes I, II ($\delta=0~\mathrm{nm}$) and III-V ($\delta=100~\mathrm{nm}$)}
\label{tab: Quality factor}
\resizebox{\columnwidth}{!}{%
\begin{tabular}{@{}ccccc@{}}
\toprule
\textbf{$\delta$}            & \textbf{Mode} & \textbf{Quality Factor ($Q_{rad}$)} & \textbf{Modulation Depth ($\%$)} & \textbf{Spectral Contrast Ratio ($\%$)} \\ \midrule
\multirow{2}{*}{$0~\mathrm{nm}$}   & I             & $2.90\times10^2$                      & $99.93$                          & 99.87                                 \\
                        & II            & $1.10\times10^2$                      & 99.78                          & 99.57                                 \\
                        \midrule
\multirow{5}{*}{$100~\mathrm{nm}$} & I & $6.14\times10^2$ & $74.51$ & $59.39$
\\
& II & $1.54\times10^2$ & $99.95$ & $99.89$
\\
& III           & $3.17\times10^3$                     & 98.50                          & 97.04                                 \\
                        & IV            & $8.96\times10^1$                       & 99.50                          & 99.00                                 \\
                        & V             & $6.38\times10^4$                    & 47.15                          & 30.85                                 \\  
                        \bottomrule
\end{tabular}%
}
\end{table*}

A simplified LC circuit based on the stable-input impedance mechanism can be employed to replicate the Fano resonance profile. Introducing a capacitor or an inductor in parallel with the LC circuit enables the coupling between a broadband bright mode and a narrowband dark mode, respectively, thus giving rise to an asymmetric Fano-like resonance~\cite{Lv2016}. Figure~\ref{fig: diff 0 100 & fit}(d) depicts the LC circuit representations for Modes I and II ($\delta = 0~\mathrm{nm}$), where a complementary capacitor and inductor were incorporated for Modes I and II, respectively. The corresponding stable-input impedances of these configurations are as follows.
\begin{equation}
    Z_{Mode~I} = -\frac{j}{\omega C_c}\frac{(\omega-\omega _{s_\circ})(\omega+\omega_{s_\circ})}{(\omega-\omega_{s_\circ}\sqrt{(C_s/C_c)+1)}(\omega+\omega_{s_\circ}\sqrt{(C_s/C_c)+1)}}
\end{equation}

\begin{equation}
    Z_{Mode~II} = \frac{j\omega L_s L_c}{(L_s + L_c)} \frac{(\omega - \omega_{s_\circ})(\omega + \omega_{s_\circ})}{(\omega - \omega_{s_\circ} / \sqrt{(L_c / L_s) + 1})(\omega + \omega_{s_\circ} / \sqrt{(L_c / L_s) + 1})}
\end{equation}
We can define transmittance as the ratio of incident power ($P_{input}$) to transmitted power ($P_{output}$). Figure~\ref{fig: diff 0 100 & fit}(e) depicts both the LC circuit fitted and calculated Modes I and II ($\delta = 0~\mathrm{nm}$), demonstrating a good match between the fitted data and the calculated result [see \textcolor{blue}{Supplementary Material} for details].\\

\begin{figure*}[htbp]
    \centering
    \includegraphics[width=1\linewidth]{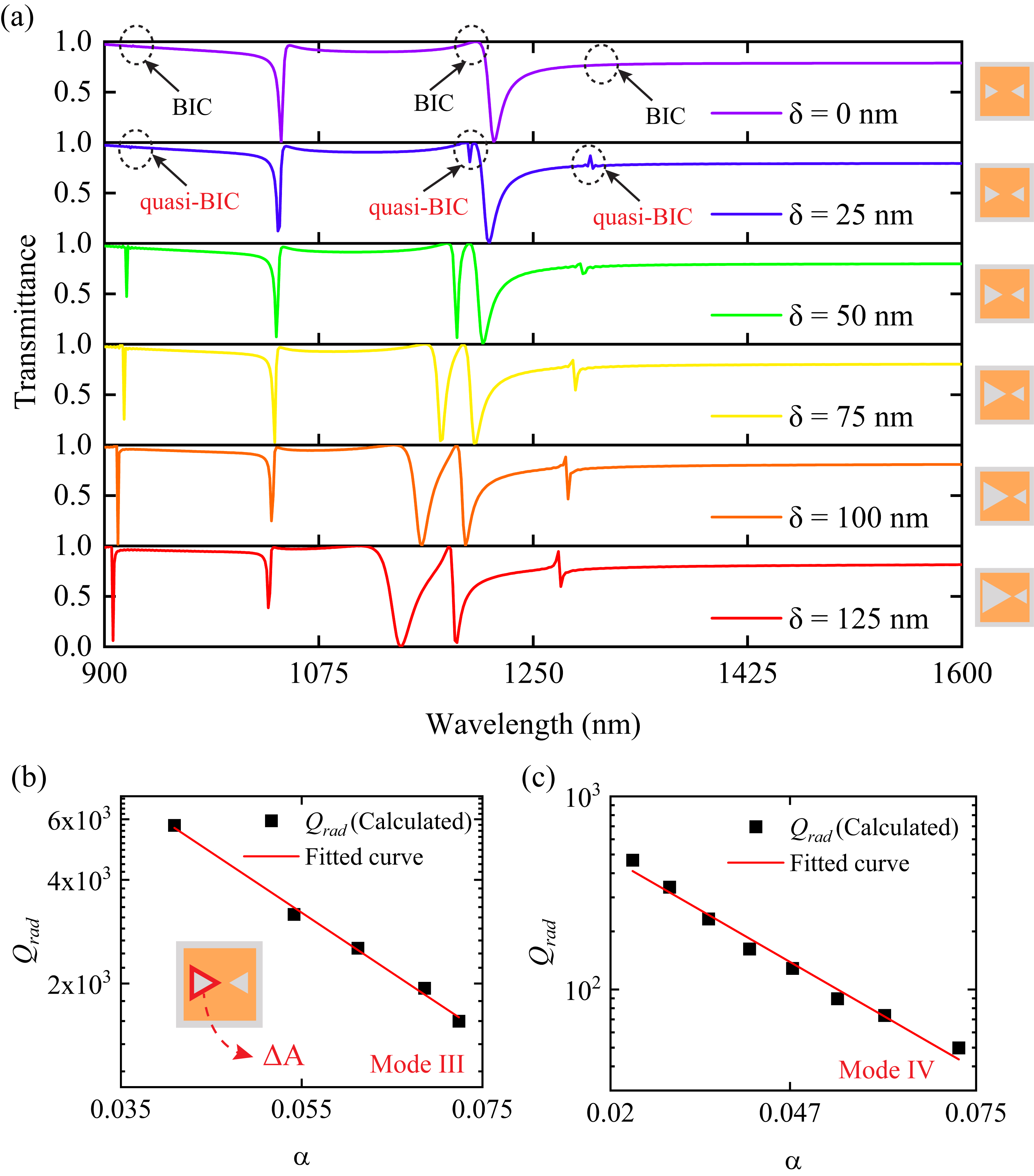}
    \caption{Transmittance spectra for different asymmetry parameters, $\delta$ = $0~\mathrm{nm}$ to $125~\mathrm{nm}$ under TM-polarized incident light (b) and (c) relationship between the Q-factor, $Q_{rad}$, and the degree of asymmetry $\alpha$ for Mode III and IV, respectively. $\alpha$ was defined by the ratio of $\Delta A$ and $A$. The solid red line represents the fitted data, indicating an inverse squared relationship between the $Q_{rad}$ and $\alpha$. Additionally, the inset of (b) highlights the magnitude of $\Delta A$, denoted by the marked red region.}
    \label{fig: side diff BIC}
\end{figure*}

Figure~\ref{fig: side diff BIC}(a) provides further insight into the formation of symmetry-breaking quasi-BIC-supported Fano resonance profiles by illustrating the transmittance spectra for $\delta = 0~\mathrm{nm}$ to $125~\mathrm{nm}$.
At $\delta = 0~\mathrm{nm}$, the transmittance spectrum exhibited three symmetry-protected BICs with a theoretically infinite Q-factor. When $\delta \neq 0~\mathrm{nm}$, the in-plane symmetry was broken, leading to the appearance of three distinct Fano profiles (Mode III to V) supported by symmetry-breaking quasi-BIC. At $\delta = 25~\mathrm{nm}$, we found these resonances at $\lambda = 921~\mathrm{nm}$, $1198~\mathrm{nm}$, and $1299~\mathrm{nm}$, while at $\delta = 125~\mathrm{nm}$, these resonances experienced a blue shift to $\lambda = 911~\mathrm{nm}$, $1159~\mathrm{nm}$, and $1278~\mathrm{nm}$ respectively. Moreover, Modes I and II also experienced the blue shift similar to the previous discussion. These blue shifts resulted from the decrease in the effective refractive index of the metasurface with the increase of the asymmetry degree $\alpha$. Here $\alpha$ was defined by $\Delta A/A$, where $\Delta A$ and $A$ represented the area of the broken part and \ce{GaP} based dielectric part, respectively. Moreover, the $Q_{rad}$ experienced a decrease with the increase of $\alpha$, resulting from the wider radiation channel, which gave rise to more energy loss through radiation. Figures~\ref{fig: side diff BIC}(b) and (c) illustrate the extracted $Q_{rad}$ as a function of $\alpha$ for Modes III and IV, which maintained the square inverse ratio law $Q_{rad}\propto(\alpha)^{-2}$. This trend satisfied the fundamental criterion for symmetry-protected BIC, indicating that the Fano resonance was governed by the BIC.\\

The underlying physical mechanism responsible for the emergence of Fano resonances can be elucidated through multipole decomposition. This method resolves the total response into distinct contributions, including the electric dipole (ED), magnetic dipole (MD), electric quadrupole (EQ), magnetic quadrupole (MQ), and toroidal dipole (TD), moments of these are mathematically defined as~\cite{Lv2024}, 
\begin{equation}
    P = \frac{1}{i\omega} \int J \, d^3r,
\end{equation}
\begin{equation}
    M = \frac{1}{2c} \int [\mathbf{r} \times \mathbf{J}] \, d^3r,
\end{equation}
\begin{equation}
    Q_{ab} = \frac{1}{2i\omega} \int \left[ r_{a}J_{b} + r_{b}J_{a} - \frac{2}{3}(\mathbf{r} \cdot \mathbf{J}) \right] d^3r,
\end{equation}
\begin{equation}
    M_{ab} = \frac{1}{3c} \int \left[ (\mathbf{r} \cdot \mathbf{J})_{a}r_{b} + r_{a}[\mathbf{r} \times \mathbf{J}]_{b}r_{a} \right] d^3r,
\end{equation}
\begin{equation}
    T = \frac{1}{10c} \int \left[ (\mathbf{r} \cdot \mathbf{J})\mathbf{r} - 2r^2\mathbf{J} \right] d^3r,
\end{equation}
where $\mathbf{J}$, $\mathbf{r}$, and $c$ represent the current density, displacement vector, and speed of light, respectively, while a and b denote the Cartesian tensor components.

\begin{figure*}[h]
    \centering
    \includegraphics[width=1\linewidth]{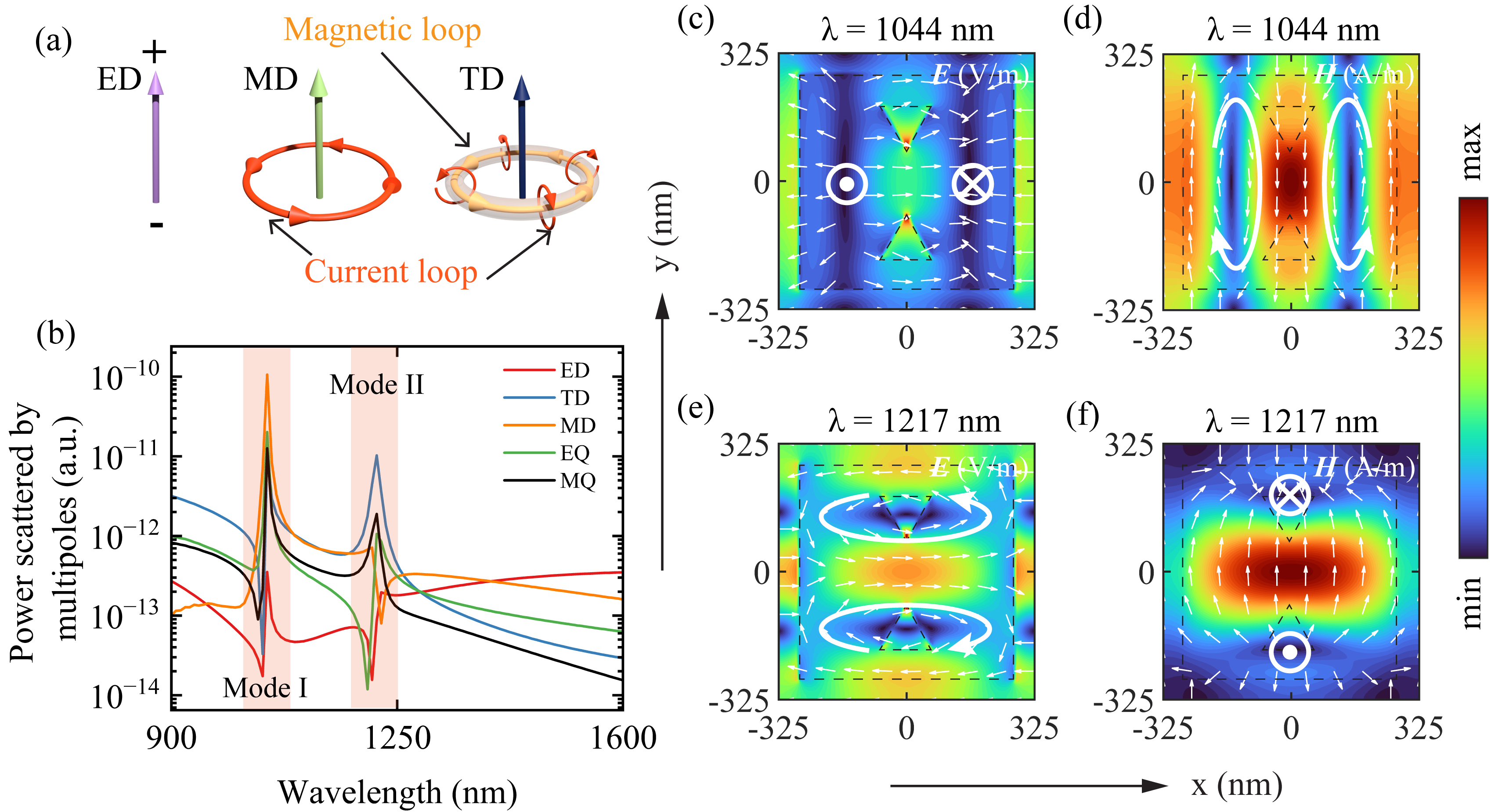}
    \caption{(a) 3D schematic representation of ED, MD, and TD modes. (b) Scattering power contributions from individual multipole moments—including ED, TD, MD, EQ, and MQ—for the proposed structure at $\delta = 0~\text{nm}$, where the red-shaded region highlights resonant modes I and II. Spatial distribution of electric and magnetic fields in the x-y plane (c-d) at $\lambda = 1044~\text{nm}$ and (e-f) at $\lambda = 1217~\text{nm}$. White arrows indicate the orientation of the corresponding field vectors.}
    \label{fig: multipole_del0}
\end{figure*}

Figure~\ref{fig: multipole_del0}(a) presents the 3D schematics of the ED, MD, and TD modes. The corresponding scattering power contributions of each multipole (ED, MD, TD, EQ, MQ) for the case of $\delta = 0~nm$ are shown in Fig.~\ref{fig: multipole_del0}(b). As observed, at a distinct resonance wavelength, a specific multipolar component dominated the scattering response. Specifically, Mode I exhibited a dominant MD contribution, while Mode II was primarily influenced by the TD response. Consequently, the electromagnetic field distribution was primarily governed by the strongest dipolar excitation. Figures~\ref{fig: multipole_del0}(c) and (d) show the spatial distributions of the electric and magnetic fields at $\lambda = 1044~\mathrm{nm}$, while Figs.~\ref{fig: multipole_del0}(e) and (f) correspond to $\lambda = 1217~\mathrm{nm}$.   
At $\lambda = 1044~\text{nm}$, the magnetic field distribution in the xy-plane featured two oppositely circulating loops, which drove an anticlockwise vortex of the electric field in the xz-plane. This configuration reflects the dominant influence of MD excitation at this resonance, oriented in the negative y direction. In contrast, at $\lambda = 1217~\text{nm}$, the electric field assumed a similar looped structure in the xy-plane, while the magnetic field exhibited a clockwise vortex in the yz-plane, revealing the dominance of TD excitation at this resonance in the positive x direction. For $\delta = 100$ nm, a pronounced TD excitation was identified at $\lambda = 911$ nm, whereas MD excitation was found to dominate at $\lambda = 1159$ nm and $\lambda = 1278$ nm [see \textcolor{blue}{Supplementary Material} for details].

\subsection{Rotational Asymmetry}
\begin{figure*}[h]
    \centering
    \includegraphics[width=1\linewidth]{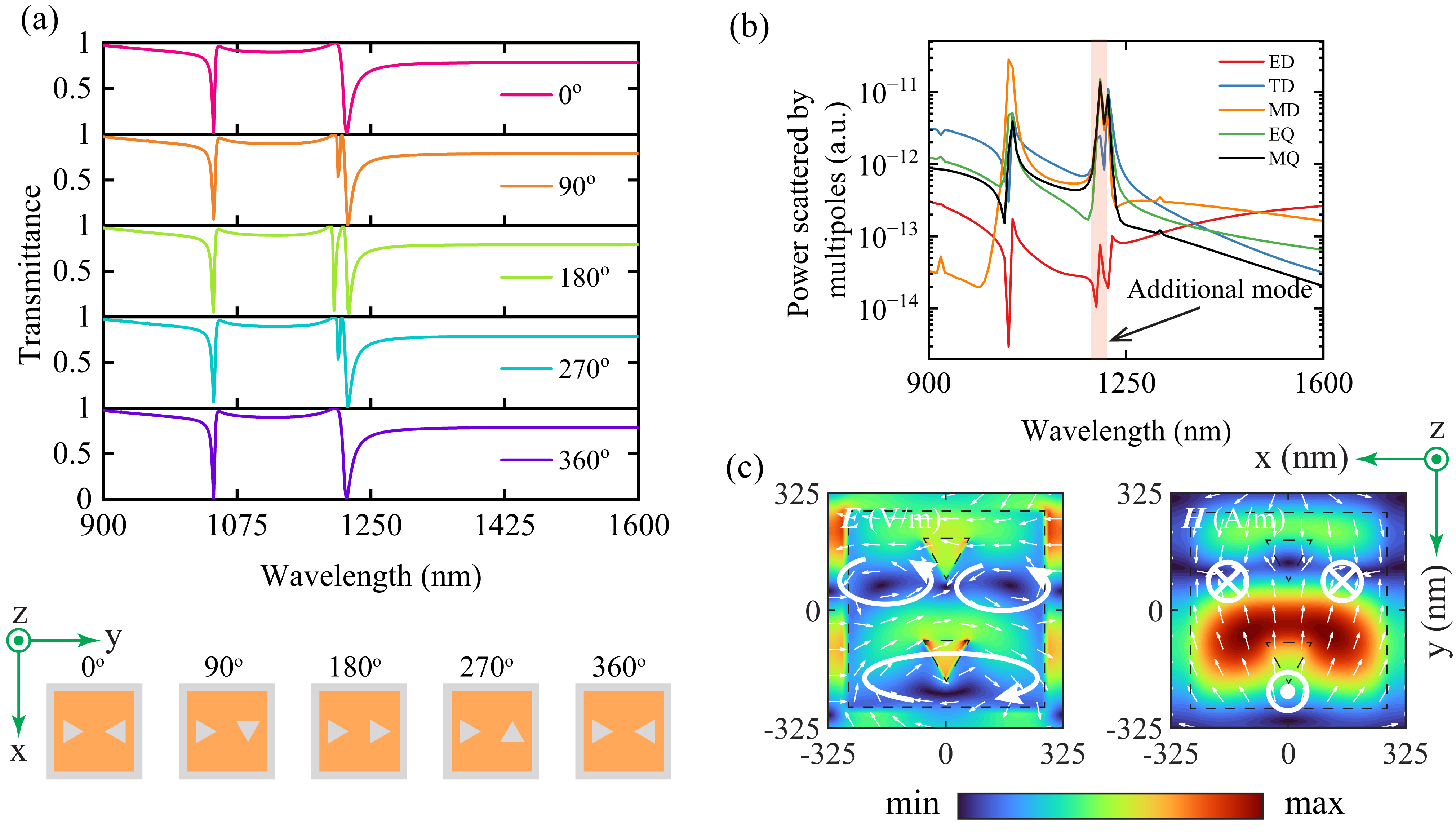}
    \caption{(a) Transmittance spectra of the proposed structure, incorporating the rotational asymmetry (one triangular nanohole rotated from $0^\circ$ to $360^\circ$ as in the figure). (b) Scattering
    power contributions from individual multipole moments at the rotation of $180^\circ$. (c) Electric and magnetic field distribution profile at $\lambda = 1201~\mathrm{nm}$ at $180^\circ$.}
    \label{fig: rotation}
\end{figure*}

The rotation of an individual triangular nanohole introduces in-plane symmetry breaking within our proposed structure. To examine this effect, we investigated the rotational dependence of a single nanohole in our structure with $\delta = 0~\mathrm{nm}$, rotating from $0^\circ$ to $360^\circ$. As shown in Fig.~\ref{fig: rotation}(a), Mode I remained at $\lambda = 1044~\mathrm{nm}$ throughout the entire rotation range, indicating its robustness to rotational asymmetry. 
In contrast, Mode II redshifted from $\lambda = 1217~\mathrm{nm}$ to $\lambda = 1221~\mathrm{nm}$ as the rotation increased from $0^\circ$ to $180^\circ$. Upon further rotation to $360^\circ$, Mode II gradually blueshifted, ultimately returning to its original resonance wavelength at $\lambda = 1217~\mathrm{nm}$. Meanwhile, an additional resonance was observed near Mode II with the rotation of this single nanohole. At a rotation angle of $90^\circ$, this resonance emerged at $\lambda = 1207~\mathrm{nm}$ with a modulation depth of $46.46~\%$. As the nanohole rotated further to $180^\circ$, the resonance exhibited a blue shift to $\lambda = 1201~\mathrm{nm}$, accompanied by a significantly enhanced modulation depth of 93.53~$\%$. The corresponding $Q_{rad}$ was $9.86 \times 10^2$ at both $90^\circ$ and $270^\circ$, but decreased to $4.40 \times 10^2$ at $180^\circ$. The observed blue shift and $Q_{rad}$ reduction with increasing asymmetry suggest that this resonance originated from a quasi-BIC. To further elucidate the origin of the additional Fano resonance, the scattering power contributions from individual multipole moments were analyzed for a rotational angle of $180^\circ$. As shown in Fig.~\ref{fig: rotation}(b), the resonance at $\lambda = 1201\mathrm{nm}$ is predominantly governed by the MD, EQ, and MQ modes. This interpretation is further supported by the electric and magnetic field distributions in Fig.~\ref{fig: rotation}(c), where two anti-clockwise and one clockwise loops in the electric field confirm the coexistence of MD and EQ characteristics. Additionally, the localized out-of-plane magnetic field excitation indicates a strong presence of MD and MQ contributions. 

\subsection{Impact of Light Polarization}
\begin{figure*}[htbp]
    \centering
    \includegraphics[width=1\linewidth]{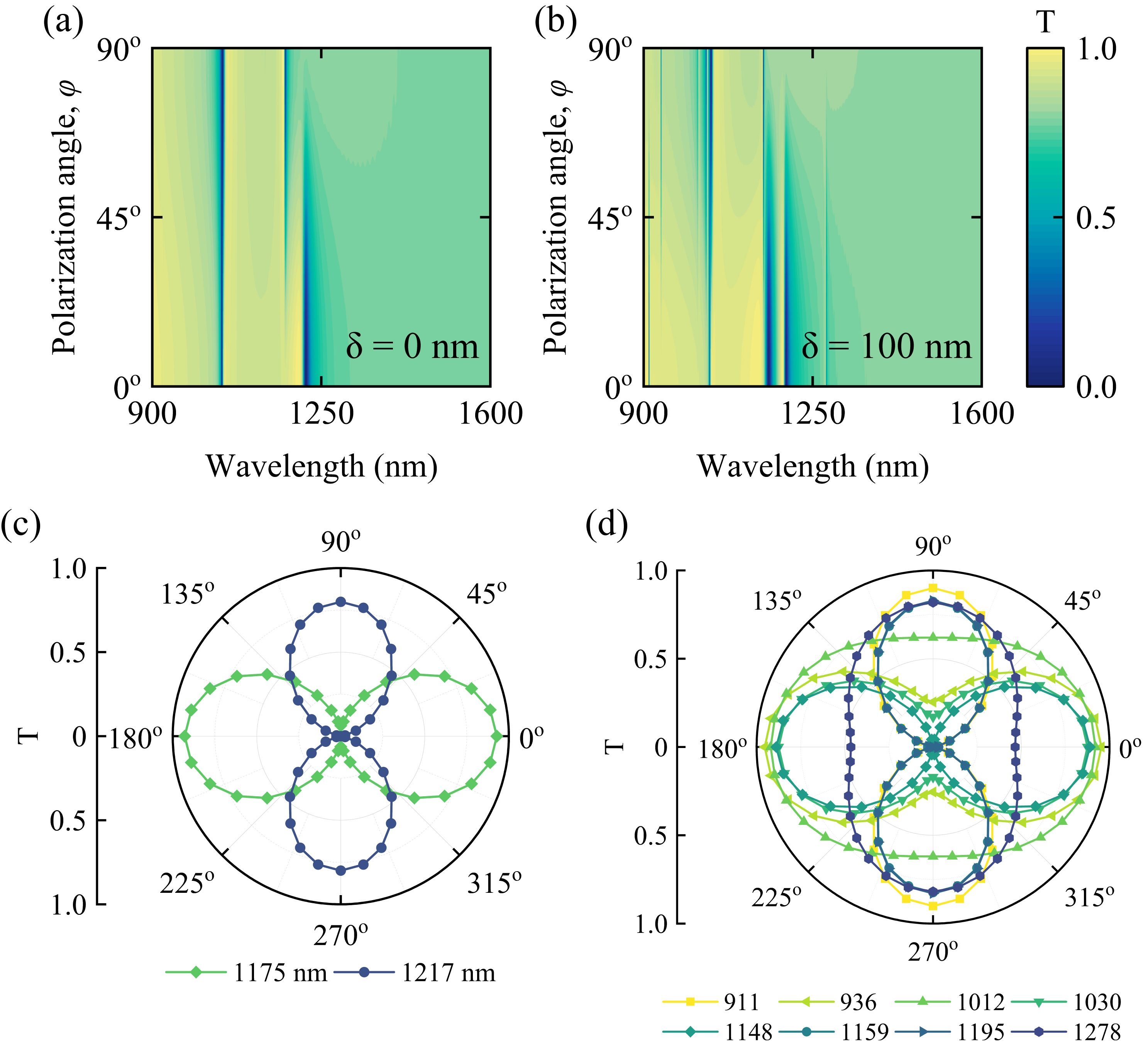}
    \caption{Transmittance spectra for (a) $\delta = 0~nm$, and (b) $\delta = 100~nm$ at polarization angle, $\varphi = 0^\circ$ to $90^\circ$. Polar plot of transmittance spectra for $\varphi = 0^\circ$ to $360^\circ$ at (c) $\lambda =$ 1175 nm and 1217 nm for $\delta =$ 0 nm and at (d) $\lambda =$ 911 nm, 936 nm, 1012 nm, 1030 nm, 1148 nm, 1159 nm, 1195 nm, and 1278 nm for $\delta =$ 100 nm.}
    \label{fig: pol angle}
\end{figure*}

Owing to the asymmetric configuration of the proposed structure,  we analyzed its optical response under varying polarization angles, $\varphi$, of the incident light. The analysis was carried out for $\varphi$ ranging from $0^\circ$ to $90^\circ$, considering structural asymmetries of $\delta = 0~\mathrm{nm}$ and $100~\mathrm{nm}$. The corresponding transmittance spectra are shown in Fig.~\ref{fig: pol angle}(a) and (b). For $\delta = 0~\mathrm{nm}$, Mode I maintained its resonance wavelength and modulation depth, though its Q-factor decreased to $1.38\times 10^2$. In contrast, Mode II exhibited a gradual reduction in modulation depth with increasing $\varphi$, and fully vanished at $\varphi = 90^\circ$. Interestingly, a new Fano profile emerged at $\lambda = 1175~\mathrm{nm}$ as $\varphi$ increased, characterized by a Q-factor of $4.34\times 10^2$ and a modulation depth of $99.96~\%$ at $\varphi = 90^\circ$. Figure~\ref{fig: pol angle}(c) illustrates the transmittance at $\lambda = $ 1175 nm, and 1217 nm for $\varphi = 0^\circ$ to $360^\circ$, presenting the obvious on-off state of the corresponding resonance modes. For $\delta = 100~\mathrm{nm}$, all Fano profiles except Mode I exhibited a gradual reduction of modulation depth with increasing $\varphi$, and totally disappeared at $\varphi = 90^\circ$. However, four new Fano profiles started to appear with the increase of $\varphi$. At $\varphi = 90^\circ$, new four Fano profiles were at $\lambda = 936~\mathrm{nm}$, $1012~\mathrm{nm}$, $1030~\mathrm{nm}$, and $1148~\mathrm{nm}$ with modulation depths of $73.06~\%$, $26.64~\%$, $80.07~\%$, and $98.21~\%$, respectively. The Q-factors of these four Fano profiles were $4.70\times10^3$, $2.95\times10^3$, $6.64\times10^2$, and $1.73\times10^3$, respectively. The transmittance at these particular wavelengths are shown in Fig.~\ref{fig: pol angle}(d) for $\varphi = 0^\circ$ to $360^\circ$ to clearly present the on-off state. The aforementioned characteristics of both structures with $\delta =0~\mathrm{nm}$ and 100 nm allow them to be utilized as an active optical switch.
\subsection{Impact of Structural Parameters}
The impact of the key structural parameters, namely, the thickness of the \ce{GaP} cuboids, t\textsubscript{GaP}, and the distance between nanoholes, d\textsubscript{nanoholes}, on the optical performance of the structure was investigated. The corresponding optical performance is discussed below. 
\subsubsection{Impact of thickness of GaP cuboids} 
\begin{figure*}
    \centering
    \includegraphics[width=1\linewidth]{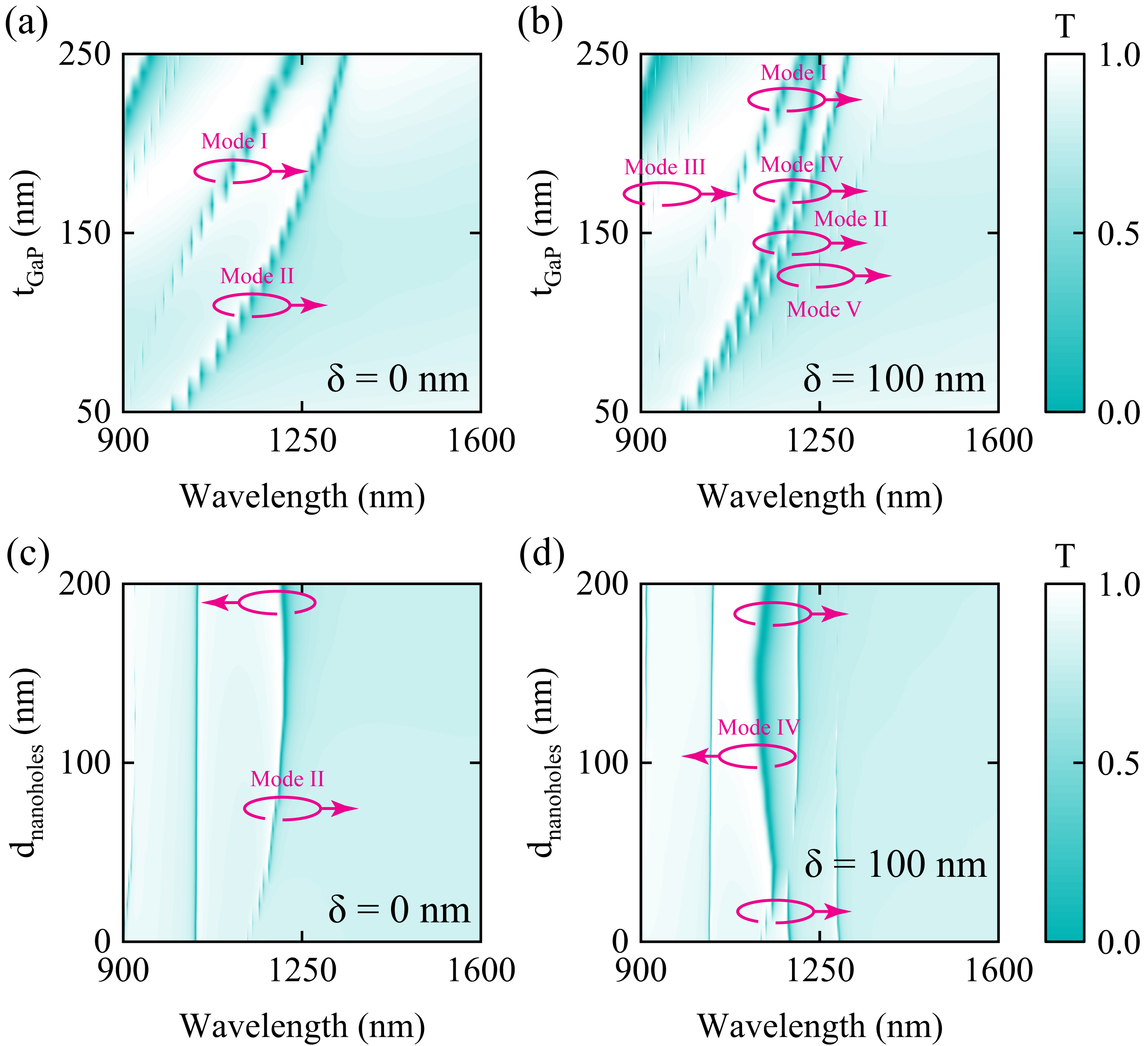}
    \caption{Transmittance spectra for varying t\textsubscript{GaP} from $50~\mathrm{nm}$ to $250~\mathrm{nm}$ at (a) $\delta = 0~\mathrm{nm}$ and (b) $\delta = 100~\mathrm{nm}$ and for varying d\textsubscript{nanoholes} from $0~\mathrm{nm}$ to $200~\mathrm{nm}$ at (c) $\delta = 0~\mathrm{nm}$ and (d) $\delta = 100~\mathrm{nm}$.}
    \label{fig: Gap thickness and distance of NH}
\end{figure*}

To examine the impact of the t\textsubscript{GaP}, we varied the t\textsubscript{GaP} from $50~\mathrm{nm}$ to $250~\mathrm{nm}$ for $\delta = 0~\mathrm{nm}$ and $100~\mathrm{nm}$, while keeping other structural parameters constant. Figures~\ref{fig: Gap thickness and distance of NH}(a) and (b) illustrate the corresponding transmittance spectra for $\delta = 0~\mathrm{nm}$ and $100~\mathrm{nm}$, respectively. For $\delta = 0~\mathrm{nm}$, the Mode I and II were located at $\lambda = 902~\mathrm{nm}$ and $\lambda = 997~\mathrm{nm}$ at t\textsubscript{GaP} = $50~\mathrm{nm}$. However, as t\textsubscript{GaP} increased, both Fano profiles exhibited a significant red shift, reaching $\lambda = 1230~\mathrm{nm}$ and $\lambda = 1332~\mathrm{nm}$  at t\textsubscript{GaP} = $250~\mathrm{nm}$, primarily due to the increased effective refractive index of the structure. Moreover, some additional resonant modes were observed for further increase of $\mathrm{t_{GaP}}>175~\mathrm{nm}$, suggesting the emergence of additional resonant pathways due to increased optical thickness. However, for $\delta = 100~\mathrm{nm}$, these additional modes appeared earlier due to introducing asymmetry in nanoholes, and a single mode (Mode III) was found for our proposed structure with t\textsubscript{GaP} = $150~\mathrm{nm}$. Despite the asymmetry ($\delta = 100~\mathrm{nm}$), the overall mode evolution with increasing t\textsubscript{GaP} followed a similar trend as the symmetric case ($\delta = 0~\mathrm{nm}$). Specifically, at t\textsubscript{GaP} = $50~\mathrm{nm}$, the Mode I and II were at $\lambda \sim 900~\mathrm{nm}$ and $\lambda = 990~\mathrm{nm}$, redshifted to $\lambda = 1214~\mathrm{nm}$ and $1302~\mathrm{nm}$ at t\textsubscript{GaP} = $250~\mathrm{nm}$. Meanwhile, Mode III emerged around t\textsubscript{GaP} of $145~\mathrm{nm}$ at $\lambda = 906~\mathrm{nm}$, and gradually shifted to $\lambda = 998~\mathrm{nm}$ at t\textsubscript{Gap} = $250~\mathrm{nm}$.  In addition, Modes IV and V, initially located at $\lambda = 980~\mathrm{nm}$ and $1036~\mathrm{nm}$ for t\textsubscript{GaP} = $50~\mathrm{nm}$, were redshifted to $\lambda = 1251~\mathrm{nm}$ and $1405~\mathrm{nm}$, respectively, as t\textsubscript{GaP} increased to $250~\mathrm{nm}$.
\subsubsection{Impact of the distance between nanoholes} The distance between the triangular nanoholes, d\textsubscript{nanoholes}, in the bow tie-shaped structure was varied to investigate the impact on optical responses for both symmetric ($\delta = 0~\mathrm{nm}$) and asymmetric ($100~\mathrm{nm}$) cases. The range of d\textsubscript{nanoholes} was considered from $0~\mathrm{nm}$ to $200~\mathrm{nm}$, and the resulting transmittance spectra are presented in Figs.~\ref{fig: Gap thickness and distance of NH}(c) and (d) for symmetric ($\delta = 0~\mathrm{nm}$) and asymmetric ($100~\mathrm{nm}$) cases, respectively. In a symmetric structure, the Mode I remained almost at the same resonance wavelength with the increase of d\textsubscript{nanoholes}. However, the Mode II experienced both redshift and blueshift, starting at $\lambda = 1143~\mathrm{nm}$ for d\textsubscript{nanoholes} = $0~\mathrm{nm}$, and ending at $\lambda = 1212~\mathrm{nm}$ for d\textsubscript{nanoholes} = $200~\mathrm{nm}$. For the asymmetric case, a similar dual-shift behavior was observed for Mode IV, which shifted from $\lambda = 1135~\mathrm{nm}$ to $\lambda = 1152~\mathrm{nm}$ across the same range of d\textsubscript{nanoholes}, while both Modes I and II remained at nearly the same resonance wavelengths. Notably, Mode III and V exhibited contradictory trends in activation: Mode III appeared at larger d\textsubscript{nanoholes}, and disappeared at smaller d\textsubscript{nanoholes}, while Mode V displayed inverse behavior.

\subsection{Sensing of \textit{Vibrio Cholerae}}
\begin{figure*}[h]
    \centering
    \includegraphics[width=1\linewidth]{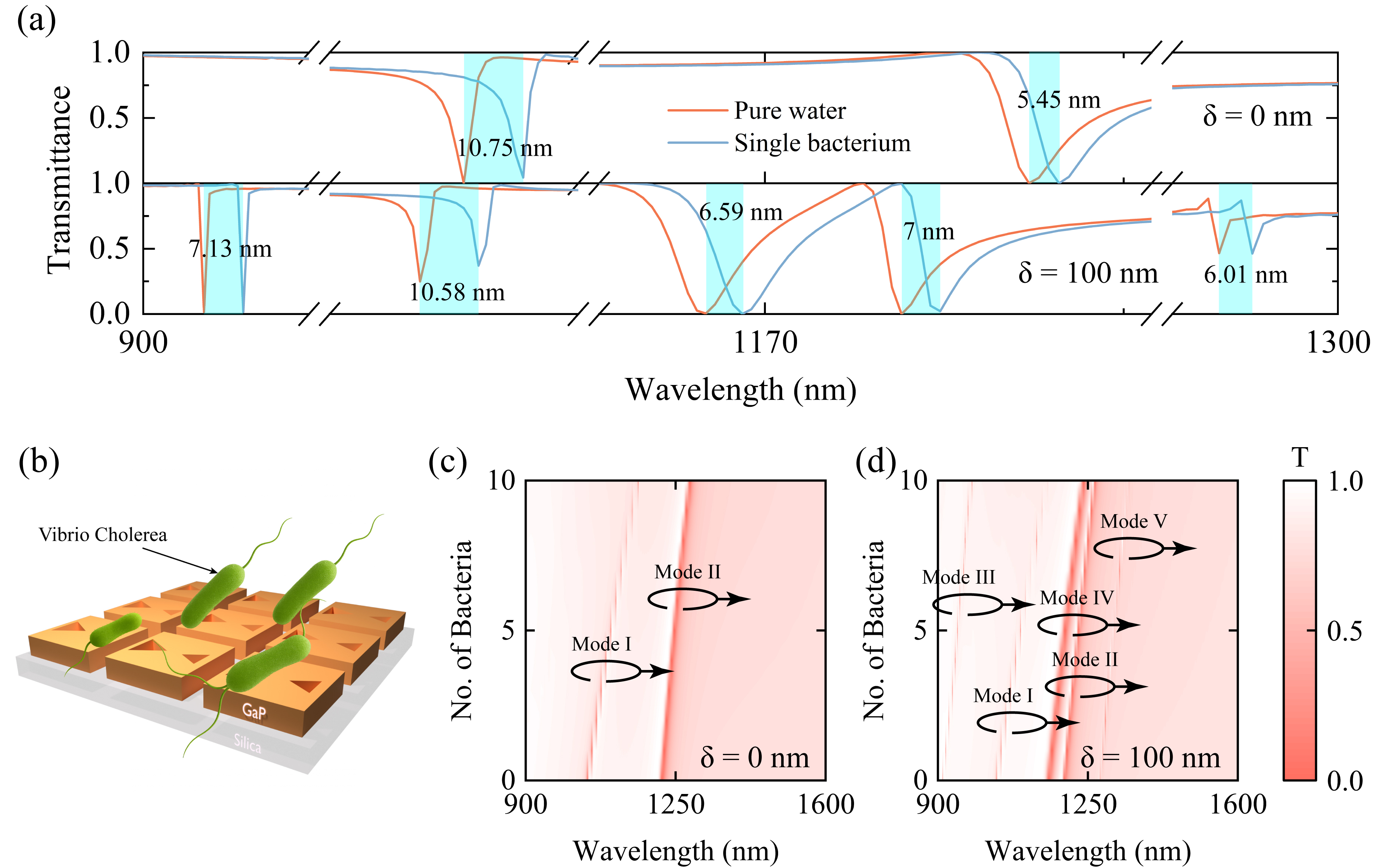}
    \caption{(a) Transmittance spectra for pure water and a single \textit{Vibrio cholerae} bacterium for $\delta = 0~\mathrm{nm}$ and $100~\mathrm{nm}$. Resonance shifts due to the presence of the bacterium are indicated by blue shading. (b) Schematic illustration of \textit{V.Cholerae} sensing by the proposed structure. (c) and (d) Transmittance spectra for 0–10 V. cholerae bacteria at $\delta = 0~\mathrm{nm}$ and $100~\mathrm{nm}$, respectively.}
    \label{fig: sensing}
\end{figure*}

\begin{figure*}[h]
    \centering
    \includegraphics[width=0.8\linewidth]{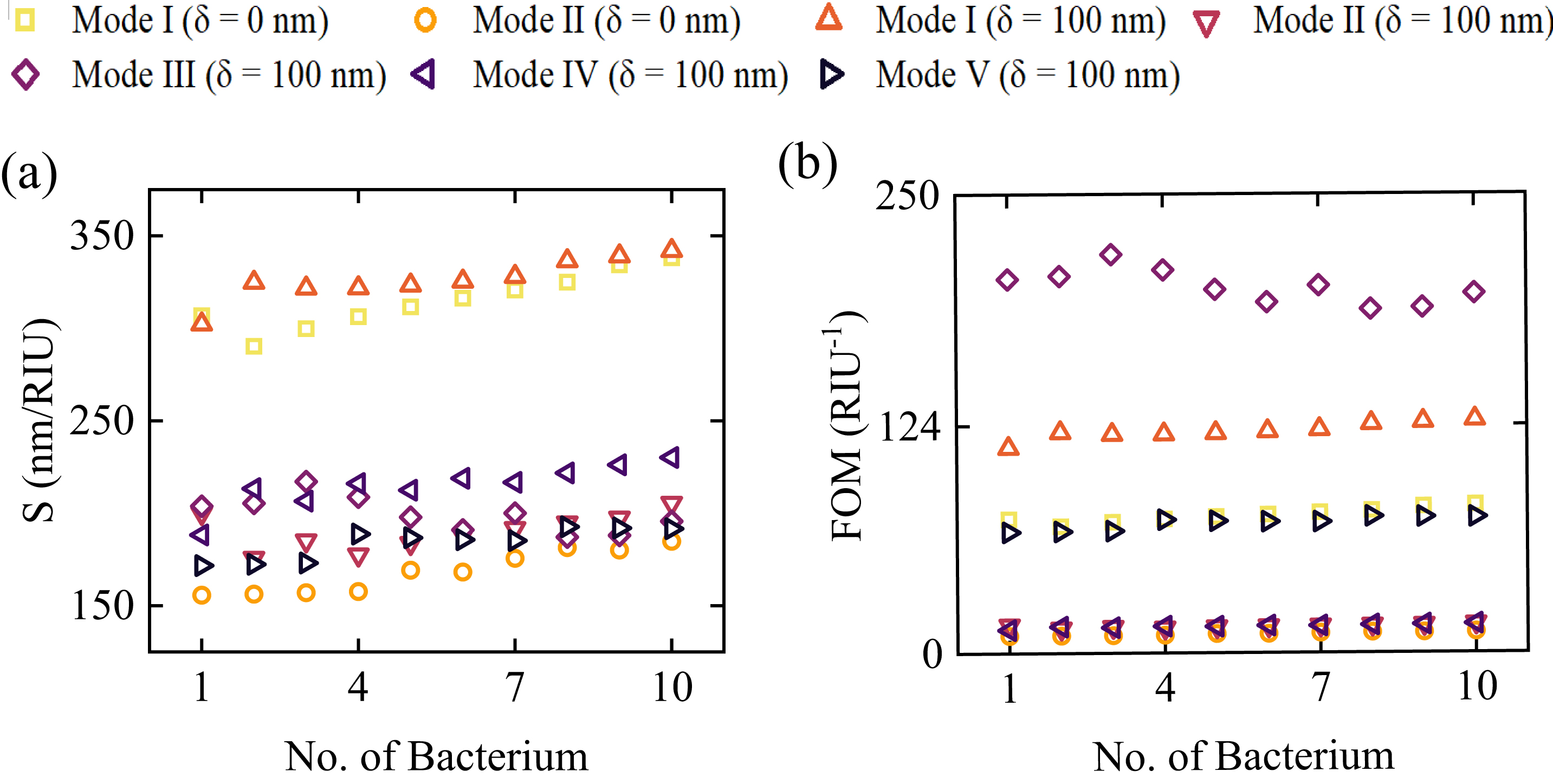}
    \caption{(a) Sensitivity (S) and (b) figure of merit (FOM) corresponding to the resonance shifts for 1–10 bacteria under both asymmetry parameters ($\delta = 0$ nm and $100$ nm.)}
    \label{fig: sensing1}
\end{figure*}

Owing to the high quality factor, $Q_{\mathrm{rad}}$ exhibited by our proposed structure, we further explored its potential for refractive index (RI) sensing. Specifically, the structure was employed to detect \textit{Vibrio cholerae}, a bacterium with an approximate refractive index of 1.365~\cite{Liu2014}. The corresponding transmittance spectra for a single bacterium for both $\delta = 0~\mathrm{nm}$ and $100~nm$ are illustrated in Fig.~\ref{fig: sensing}(a), where the associated resonance wavelength shifts are also marked. For both cases, the most prominent resonance shift was associated with Mode I, which was 10.75 nm and 10.58 nm for $\delta = 0~\mathrm{nm}$ and $100~\mathrm{nm}$, respectively; however, the modulation depth was lower (74.51~\%) for $\delta = 100~\mathrm{nm}$.  Interestingly, Mode III at $\delta = 100~\mathrm{nm}$ offered a strong trade-off, exhibiting both a high $Q_{\mathrm{rad}}$ and significant modulation depth, alongside a resonance shift of 7.13 nm. Figures~\ref{fig: sensing}(c) and (d) depict the transmittance spectra for an increasing number of \textit{V. cholerae} bacteria (0 to 10) for $\delta = 0~\mathrm{nm}$, and $100~\mathrm{nm}$, respectively. In both cases, the resonance wavelength exhibited a redshift, attributed to the increase in the effective refractive index introduced by additional bacteria cells. The corresponding performance parameters, including sensitivity (S) and figure of merit (FOM) are shown in Figs.~\ref{fig: sensing1}(a) and (b), respectively, which were calculated by, 
\begin{equation}
    \mathrm{S} = \frac{\Delta\lambda}{\Delta n},
\end{equation}
\begin{equation}
    \mathrm{FOM} = \frac{\mathrm{S}}{\mathrm{FWHM}}.
\end{equation}
Here, $\Delta\lambda$, $\Delta n$, and $\mathrm{FWHM}$ represent the resonance wavelength shift, the change in refractive index, and the full width at half maximum of the resonance, respectively. As previously discussed, Mode I exhibited the most prominent resonance wavelength shifts for both $\delta = 0~\mathrm{nm}$ and $\delta = 100~\mathrm{nm}$, which was directly reflected in its sensitivity. For single bacterium detection, Mode I at $\delta = 0~\mathrm{nm}$ yielded the highest sensitivity among all modes, with a value of 306.85 nm/RIU. A comparable sensitivity of 302.28 nm/RIU was observed for Mode I at $\delta = 100~\mathrm{nm}$. In both symmetric and asymmetric cases, Mode I exhibited a consistent increase in sensitivity with the number of bacteria, reaching approximately 342~nm/RIU when ten bacteria were present.\\
Since the FOM is strongly related to the quality factor, Mode III ($\delta = 100~\mathrm{nm}$) achieved significantly higher FOM values than the other modes, maintaining an average around 200 RIU\textsuperscript{-1}, peaking at 217.14 RIU\textsuperscript{-1} for three bacteria. Interestingly, despite its superior sensitivity, Mode I at $\delta = 100~\mathrm{nm}$ showed relatively lower FOM values, fluctuating around 124~RIU\textsuperscript{-1}. Consequently, Mode III ($\delta=100~\mathrm{nm}$) outperformed in terms of FOM, indicating its potential advantage for applications requiring high spectral resolution and detection precision.

\section{Comparative Analysis}
Table~\ref{tab: comparison} summarizes the key performance metrics of previously reported all-dielectric metastructures exhibiting Fano resonances, highlighting the compelling performance balance achieved by the proposed bowtie-shaped nanohole-based cuboid metasurface.
Zhang \textit{et al.} employed a lucky knot structure using Si and \ce{BaF2} materials in the mid-infrared region, yielding a relatively low Q-factor of $5.20\times10^2$ and a FOM of only 32.7 RIU\textsuperscript{-1}, although the authors reported a high sensitivity (S) of 986 nm/RIU~\cite{Zhang2021}. Li \textit{et al.} designed a structure with a nanocube with two square air holes operating in the 1100--1600 nm range, which achieved an ultrahigh Q-factor of $1.54\times10^5$ and nearly 100~\% modulation depth by exploiting TD and MD resonances~\cite{Li2021}. However, the derived FOM of 389 RIU\textsuperscript{-1}, and the sensitivity 287.5 nm/RIU remained moderate. In another work, cuboid tetramer clusters reported by Yu \textit{et al.} demonstrated strong multipolar interactions (TD, MQ) and yielded a FOM of 910 RIU\textsuperscript{-1}, with a Q-factor of $1.13\times10^4$ and S = 182 nm/RIU, but again operated over a narrower spectral window (1080--1200 nm). Lv \textit{et al.} pushed the sensing performance further with a four-pillar with GaP, showing an extremely high S of 488.99 nm/RIU and an exceptional FOM of $2.51\times10^5$ RIU\textsuperscript{-1}, facilitated by a combination of EQ, MQ, and MD resonances. Pang \textit{et al.} introduced a nanoblock tetramer cluster incorporating square defects, achieving a Q-factor of $1.66\times10^4$ and nearly 100~\% modulation depth. Their structure supported TD and MQ excitations, delivering an excellent sensitivity of 256 nm/RIU and a high FOM of 2519.7 RIU\textsuperscript{-1}. Li \textit{et al.} proposed an I-shaped bar and $\Phi$-shaped disk configuration with structural asymmetry based on distance variation. The design achieved a Q-factor of $1.77\times10^4$ and high sensitivity S = 784.8 nm/RIU through TD and MD excitation.\\
In contrast, our work presents a bowtie-shaped nanohole-based cuboid comprised of GaP, designed for the near-infrared region, a range more compatible with on-chip integration and bio-sensing. The incorporated geometric asymmetry enables the excitation of both TD and MD modes, peaking Q-factor of $6.38\times10^4$, near-unity modulation depth, and an appreciable sensitivity of 342 nm/RIU. The corresponding FOM of 217.14 RIU\textsuperscript{-1} surpassed most of recent works while maintaining a compact and fabrication-friendly configuration.

\begin{table*}[]
\centering
\renewcommand{\arraystretch}{1.8} 
{\fontsize{5.5pt}{9pt}\selectfont 
\caption{Comparison of all-dielectric metastructures exhibiting Fano resonances, highlighting key performance metrics. The last row corresponds to the proposed structure.}
\label{tab: comparison}
\begin{tabular}{@{}P{1.8cm} P{0.6cm} P{1.3cm} P{1.15cm} P{0.7cm} P{1cm} P{0.8cm} P{0.8cm} P{0.8cm} P{0.5cm}@{}}
\toprule
\textbf{Structure Type} & \textbf{Material} & \textbf{Operating Wavelength (nm)} & \textbf{Asymmetry Parameter} & \textbf{Quality Factor} & \textbf{Modulation Depth} & \textbf{Excitation Mode} & \textbf{S (nm/RIU)} & \textbf{FOM (RIU\textsuperscript{-1})} & \textbf{Ref} \\ \midrule
Lucky knot & Si, \ce{BaF2} & 6500–8000 & – & 5.20$\times10^2$ & – & – & 986 & 32.7 & \cite{Zhang2021} \\
Nanocube with two square air holes & Si, \ce{SiO2} & 1100–1600 & Air hole dimension & 1.54$\times10^5$ & nearly 100\% & TD, MD & 287.5 & 389 & \cite{Li2021} \\
Cuboid tetramer clusters with symmetric structural parameters & – & 1080–1200 & Permittivity & 1.13$\times10^4$ & nearly 100\% & TD, MQ & 182 & 910 & \cite{Yu2022} \\
Four cylindrical pillars & GaP, \ce{MgF2} & 1030–1180 & Placement of two pillars & 1.47$\times10^4$ & – & EQ, MQ, MD & 488.99 & 2.51$\times10^5$ & \cite{Lv2024} \\
Nanoblock tetramer clusters & Si, \ce{SiO2} & 1000–1350 & Square defects & 1.66$\times10^4$ & nearly 100\% & TD, MQ & 256 & 2519.7 & \cite{Pang2024} \\
I-shaped bar and $\phi$-shaped disk & Si & 1100–1300 & Distance between bar and neighboring disks & 1.77$\times10^4$ & – & TD, MD & 784.8 & – & \cite{Li2021} \\
Bowtie-shaped nanohole-based cuboid & GaP, \ce{SiO2} & 900–1600 & Dimension of triangular nanohole & 6.38$\times10^4$ & nearly 100\% & TD, MD & 342 & 217.14 & This Work \\
\bottomrule
\end{tabular}
} 
\end{table*}
\begin{figure}[h]
    \centering
    \includegraphics[width=1\linewidth]{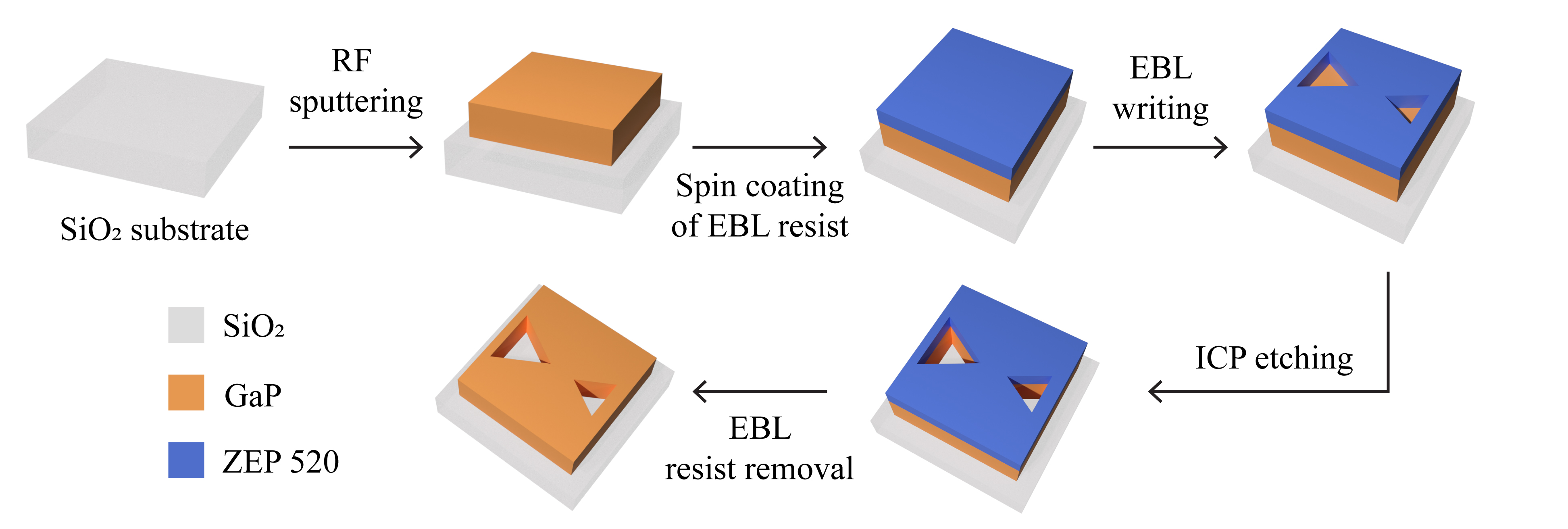}
    \caption{Suggested fabrication techniques of our proposed metasurface structure with possible steps from \ce{SiO2} substrate to final structure.}
    \label{fig: fab}
\end{figure}

\section{Suggested Fabrication Techniques}
Although our study was conducted through numerical simulation, previous studies have corroborated the experimental feasibility of fabricating the proposed structure.  The fabrication sequence, from the initial \ce{SiO2} substrate to the completed device, is schematically presented in Fig.~\ref{fig: fab}. The \ce{GaP} cuboid layer can be deposited through radio-frequency (RF) sputtering~\cite{Tilmann2020}, a well-established technique for achieving uniform and high-quality semiconductor thin films. To define the bow-tie-shaped nanohole patterns, electron beam lithography (EBL) can be employed. This process starts with the spin-coating of an EBL-sensitive resist layer on top of the \ce{GaP} surface. After patterning the resist using high-resolution EBL writing, the design can be transferred into the \ce{GaP} layer via inductively coupled plasma (ICP) etching~\cite{Li2021}. Subsequent removal of the residual resist will complete the fabrication, yielding the final device structure with the designed nanophotonic features.

\section{Conclusions}
We proposed a highly versatile, ingenious all-dielectric metasurface, harnessing precisely engineered structural asymmetry to produce multiple high-Q quasi-BIC Fano resonances. The physical origins of these modes were clarified through multipolar decomposition, revealing dominant MD and TD contributions. To further validate the resonance features, LC circuit modeling and theoretical Fano profile fitting were employed, while the BIC nature was confirmed through the square inverse law. The proposed structure offered a remarkable combination of spectral sensitivity, modulation depth, and design simplicity with Q-factor approaching $10^4$, pronounced polarization selectivity, and notable refractive index sensing performance, where sensitivity of 342 nm/RIU and FOM of 217.14 RIU$^{-1}$ were achieved for \textit{Vibrio Cholerae}. Furthermore, its CMOS compatibility and tunable optical response significantly broadened its applicability, reflecting it as a powerful candidate for next-generation integrated photonic technologies, including high-performance biosensing, active optical modulation, nonlinear light–matter interactions, and beyond.

\section*{CRediT authorship contribution statement}
\textbf{Soikot Sarkar}: Conceptualization, Methodology, Visualization, Software, Investigation, Writing – original draft, Writing – review \& editing. 
\textbf{Ahmed Zubair}: Conceptualization, Methodology, Visualization, Resources, Supervision, Writing – original draft, Writing – review \& editing.

\section*{Declaration of Competing Interest}
The authors declare no conflicts of interest.

\section*{Data availability}
Data underlying the results presented in this paper are not publicly available at this time but may be obtained from the authors upon reasonable request.

\section*{Acknowledgement}
S. Sarkar and A. Zubair acknowledge the Department of Electrical and Electronic Engineering, Bangladesh University of Engineering and Technology, for providing access to Ansys Lumerical software and essential computational support. 

\bibliographystyle{elsarticle-num} 
\bibliography{sample}

\begin{thebibliography}{10}
\expandafter\ifx\csname url\endcsname\relax
  \def\url#1{\texttt{#1}}\fi
\expandafter\ifx\csname urlprefix\endcsname\relax\def\urlprefix{URL }\fi
\expandafter\ifx\csname href\endcsname\relax
  \def\href#1#2{#2} \def\path#1{#1}\fi

\bibitem{Shaltout2019}
A.~M. Shaltout, V.~M. Shalaev, M.~L. Brongersma, Spatiotemporal light control with active metasurfaces, Science 364~(6441) (2019) eaat3100.

\bibitem{Pertsch2023}
T.~Pertsch, S.~Xiao, A.~Majumdar, G.~Li, Optical metasurfaces: fundamentals and applications, Photonics Research 11~(5) (2023) OMFA1--OMFA3.

\bibitem{Dip2025JMCC}
D.~Sarker, A.~Zubair, Advances in hyperbolic metamaterial sensors: a comprehensive review, J. Mater. Chem. C 13 (2025) 19106--19124.

\bibitem{Sarkar2024}
S.~Sarkar, D.~Sarker, A.~Zubair, A polarization tunable incident angle tolerant dielectric metasurface-based color filter, Materials Advances 5~(18) (2024) 7455--7466.

\bibitem{Nakti2023}
P.~P. Nakti, D.~Sarker, M.~I. Tahmid, A.~Zubair, Ultra-broadband near-perfect metamaterial absorber for photovoltaic applications, Nanoscale Advances 5~(24) (2023) 6858--6869.

\bibitem{Kuznetsov2024}
A.~I. Kuznetsov, M.~L. Brongersma, J.~Yao, M.~K. Chen, U.~Levy, D.~P. Tsai, N.~I. Zheludev, A.~Faraon, A.~Arbabi, N.~Yu, D.~Chanda, K.~B. Crozier, A.~V. Kildishev, H.~Wang, J.~K.~W. Yang, J.~G. Valentine, P.~Genevet, J.~A. Fan, O.~D. Miller, A.~Majumdar, J.~E. Fr{\"o}ch, D.~Brady, F.~Heide, A.~Veeraraghavan, N.~Engheta, A.~Al{\`u}, A.~Polman, H.~A. Atwater, P.~Thureja, R.~Paniagua-Dominguez, S.~T. Ha, A.~I. Barreda, J.~A. Schuller, I.~Staude, G.~Grinblat, Y.~Kivshar, S.~Peana, S.~F. Yelin, A.~Senichev, V.~M. Shalaev, S.~Saha, A.~Boltasseva, J.~Rho, D.~K. Oh, J.~Kim, J.~Park, R.~Devlin, R.~A. Pala, Roadmap for optical metasurfaces, ACS Photonics 11~(3) (2024) 816--865.

\bibitem{Koshelev2019}
K.~Koshelev, A.~Bogdanov, Y.~Kivshar, Meta-optics and bound states in the continuum, Science Bulletin 64~(12) (2019) 836--842.

\bibitem{Islam2025}
O.~Islam, D.~Sarker, K.~B. M.~S. Mahmood, J.~Debnath, A.~Zubair, Bird's eye inspired hyperuniform disordered tio2 meta-atom based high-efficiency metalens, Nanoscale Advances 7~(4) (2025) 1134--1142.

\bibitem{Chen2024}
H.~Chen, X.~Fan, W.~Fang, B.~Zhang, S.~Cao, Q.~Sun, D.~Wang, H.~Niu, C.~Li, X.~Wei, C.~Bai, S.~Kumar, High-q fano resonances in all-dielectric metastructures for enhanced optical biosensing applications, Biomedical Optics Express 15~(1) (2024) 294--305.

\bibitem{Kang2023}
M.~Kang, T.~Liu, C.~T. Chan, M.~Xiao, Applications of bound states in the continuum in photonics, Nature Reviews Physics 5~(11) (2023) 659--678.

\bibitem{Yi2025}
Y.~Yi, Q.~Song, L.~Jiang, Z.~Yi, Y.~Yi, M.~Long, Simultaneously tuned ultra-high q and multimodal resonance in magnetic dipole bic systems governed by coherent perfect reflection principle, Optics {\&} Laser Technology 183 (2025) 112372.

\bibitem{Jafari2024}
E.~Jafari, M.~A. Mansouri-Birjandi, A.~Tavousi, High-performance plasmonic metasurface sensor by triangular nano-structures, Optics Continuum 3~(1) (2024) 78--93.

\bibitem{Lv2024}
J.~Lv, Y.~Ren, D.~Wang, J.~Wang, X.~Lu, Y.~Yu, W.~Li, Q.~Liu, X.~Xu, W.~Liu, P.~K. Chu, C.~Liu, Optical switching with high-q fano resonance of all-dielectric metasurface governed by bound states in the continuum, Optics Express 32~(16) (2024) 28334--28347.

\bibitem{Yang2024}
X.~Yang, H.~Xu, H.~Xu, M.~Li, H.~Yu, Y.~Cheng, Z.~Chen, Terahertz refractive index sensor based on dual plasmon-induced transparency in a graphene metasurface, Physica Scripta 99~(5) (2024) 055518.

\bibitem{Sarker2021}
D.~Sarker, P.~P. Nakti, M.~I. Tahmid, M.~A.~Z. Mamun, A.~Zubair, Terahertz polarizer based on tunable surface plasmon in graphene nanoribbon, Optics Express 29~(26) (2021) 42713--42725.

\bibitem{Liang2024}
Y.~Liang, D.~P. Tsai, Y.~Kivshar, From local to nonlocal high-{\$}q{\$} plasmonic metasurfaces, Physical Review Letters 133~(5) (2024) 053801.

\bibitem{Wang2023}
D.~Wang, X.~Fan, W.~Fang, H.~Niu, J.~Tao, C.~Li, X.~Wei, Q.~Sun, H.~Chen, H.~Zhao, Y.~Yin, W.~Zhang, C.~Bai, S.~Kumar, Excitation of multiple fano resonances on all-dielectric nanoparticle arrays, Optics Express 31~(6) (2023) 10805--10819.

\bibitem{Wang2021}
J.~Wang, J.~K{\"u}hne, T.~Karamanos, C.~Rockstuhl, S.~A. Maier, A.~Tittl, All-dielectric crescent metasurface sensor driven by bound states in the continuum, Advanced Functional Materials 31~(46) (2021) 2104652.

\bibitem{Song2023}
F.~Song, B.~Xiao, J.~Qin, High-q multiple fano resonances with near-unity modulation depth governed by nonradiative modes in all-dielectric terahertz metasurfaces, Optics Express 31~(3) (2023) 4932--4941.

\bibitem{Xu2021}
J.~Xu, H.~Fan, Q.~Dai, H.~Liu, S.~Lan, Toroidal dipole response in the individual silicon hollow cylinder under radially polarized beam excitation, Journal of Physics D: Applied Physics 54~(21) (2021) 215102.

\bibitem{Li2022}
M.~Li, Q.~Ma, A.~Luo, W.~Hong, Multiple toroidal dipole symmetry-protected bound states in the continuum in all-dielectric metasurfaces, Optics {\&} Laser Technology 154 (2022) 108252.

\bibitem{Li2021}
H.~Li, S.~Yu, L.~Yang, T.~Zhao, High q-factor multi-fano resonances in all-dielectric double square hollow metamaterials, Optics {\&} Laser Technology 140 (2021) 107072.

\bibitem{Sarker2024_PCCP}
D.~Sarker, A.~Zubair, Titanium nitride-based hyperbolic metamaterial for near-infrared ultrasensitive sensing of microbes, Physical Chemistry Chemical Physics 26~(13) (2024) 10273--10283.

\bibitem{Xie2023}
S.~Xie, S.~Sun, Z.~Li, J.~Yang, W.~Shen, X.~Guan, Manipulation of the multiple bound states in the continuum and slow light effect in the all-dielectric metasurface, Journal of Physics D: Applied Physics 56~(40) (2023) 405109.

\bibitem{Chen2025}
C.~Chen, R.~He, J.~Guo, High-quality factor resonance enabled by the weak coupling between te and tm resonance modes in a two-dimensional asymmetric metasurface and its application for slow light, Optics Express 33~(4) (2025) 8712--8725.

\bibitem{Feng2023}
J.~Feng, L.-S. Wu, J.-F. Mao, Switchable broadband/narrowband absorber based on a hybrid metasurface of graphene and metal structures, Optics Express 31~(8) (2023) 12220--12231.

\bibitem{Ye2022}
Y.~Ye, S.~Yu, H.~Li, Z.~Gao, L.~Yang, T.~Zhao, Triple fano resonances metasurface and its extension for multi-channel ultra-narrow band absorber, Results in Physics 42 (2022) 106025.

\bibitem{Wang2024}
T.~Wang, S.~Liu, J.~Zhang, L.~Xu, M.~Yang, B.~Han, D.~Ma, S.~Jiang, Q.~Jiao, X.~Tan, Highly sensitive polarization-tunable fano resonant metasurface excited by bics for refractive index detection, Results in Physics 58 (2024) 107451.

\bibitem{Wang2024_AO}
T.~Wang, W.~Fang, H.~Guo, J.~Pang, X.~Fan, C.~Li, X.~Wei, S.~Kumar, High q-factor fano resonances based on an all-dielectric metasurface for high-performance refractive index sensing, Applied Optics 63~(24) (2024) 6322--6330.

\bibitem{Yang2021}
L.~Yang, S.~Yu, H.~Li, T.~Zhao, Multiple fano resonances excitation on all-dielectric nanohole arrays metasurfaces, Optics Express 29~(10) (2021) 14905--14916.

\bibitem{Bhowmik2024}
B.~K. Bhowmik, K.~M. Rohith, P.~Duhan, G.~Kumar, Excitation of high-quality quasi-bic toroidal mode in a lattice perturbed terahertz metasurface, Applied Physics Letters 125~(16) (2024) 161701.

\bibitem{Liu2024_RSC}
J.~Liu, H.~Dai, J.~Ju, K.~Cheng, A triple fano resonance si--graphene metasurface for multi-channel tunable ultra-narrow band sensing, Physical Chemistry Chemical Physics 26~(12) (2024) 9462--9474.

\bibitem{Dong2025}
H.~Dong, Y.~He, H.~Zhu, Y.~Chen, Y.~Zhang, J.~Wang, Polarization-selective high-sensitivity fano resonance in all-dielectric metasurface based on quasi-bound states in the continuum, Optics Communications 592 (2025) 132273.

\bibitem{Zhao2024}
C.~Zhao, Y.~Huo, T.~Liu, Z.~Liao, C.~Xu, T.~Zhang, All-dielectric metasurface with multiple fano resonances supporting high-performance refractive index sensing, Journal of the Optical Society of America B 41~(1) (2024) 36--45.

\bibitem{Sun2024}
F.~Sun, X.~Fan, W.~Fang, J.~Zhao, W.~Xiao, C.~Li, X.~Wei, J.~Tao, Y.~Wang, S.~Kumar, Multiple toroidal dipole fano resonances from quasi-bound states in the continuum in an all-dielectric metasurface, Optics Express 32~(10) (2024) 18087--18098.

\bibitem{Pang2024}
J.~Pang, W.~Fang, X.~Fan, Q.~Chen, H.~Guo, T.~Wang, X.~Wei, C.~Bai, S.~Kumar, Polarization-independent tetramer metastructure with multi-fano resonances governed by quasi-bound states in the continuum, Optics Express 32~(18) (2024) 31905--31919.

\bibitem{Yu2022}
S.~Yu, Y.~Wang, Z.~Gao, H.~Li, S.~Song, J.~Yu, T.~Zhao, Dual-band polarization-insensitive toroidal dipole quasi-bound states in the continuum in a permittivity-asymmetric all-dielectric meta-surface, Optics Express 30~(3) (2022) 4084--4095.

\bibitem{Bond1965}
W.~L. Bond, Measurement of the refractive indices of several crystals, Journal of Applied Physics 36~(5) (1965) 1674--1677.

\bibitem{Palik1998}
E.~D. Palik, Handbook of Optical Constants of Solids, no. v. 3 in Academic Press handbook series, Elsevier Science, 1998.

\bibitem{Lv2016}
B.~Lv, R.~Li, J.~Fu, Q.~Wu, K.~Zhang, W.~Chen, Z.~Wang, R.~Ma, Analysis and modeling of fano resonances using equivalent circuit elements, Scientific Reports 6~(1) (2016) 31884.

\bibitem{Liu2014}
P.~Y. Liu, L.~K. Chin, W.~Ser, T.~C. Ayi, P.~H. Yap, T.~Bourouina, Y.~Leprince-Wang, An optofluidic imaging system to measure the biophysical signature of single waterborne bacteria, Lab on a Chip 14~(21) (2014) 4237--4243.

\bibitem{Zhang2021}
Y.~Zhang, Z.~Liang, D.~Meng, Z.~Qin, Y.~Fan, X.~Shi, D.~R. Smith, E.~Hou, All-dielectric refractive index sensor based on fano resonance with high sensitivity in the mid-infrared region, Results in Physics 24 (2021) 104129.

\bibitem{Tilmann2020}
B.~Tilmann, G.~Grinblat, R.~Bert{\'e}, M.~{\"O}zcan, V.~F. Kunzelmann, B.~Nickel, I.~D. Sharp, E.~Cort{\'e}s, S.~A. Maier, Y.~Li, Nanostructured amorphous gallium phosphide on silica for nonlinear and ultrafast nanophotonics, Nanoscale Horizons 5~(11) (2020) 1500--1508.

\end{thebibliography}

\end{document}